\documentclass[11pt]{article}
\usepackage{moriond,epsfig}
\usepackage{amsmath, amssymb}

\def\be{\begin{equation}}
\def\ee{\end{equation}}
\def\bea{\begin{eqnarray}}
\def\eea{\end{eqnarray}}

\begin{document}
\vspace*{4cm}
\title{Electroweak Baryogenesis in a Two-Higgs Doublet Model}

\author{Stephan J.~Huber}

\address{CERN, Theory Division, CH-1211 Geneva 23, Switzerland}

\maketitle\abstracts{
Electroweak baryogenesis fails in the SM because of too small CP violation
and the lack of a strong first-order phase transition. It has been shown
that supersymmetric models allow for successful baryogenesis, where the Higgsinos
play an important role in the transport processes that generate the asymmetry.
I will demonstrate that also non-supersymmetric models can provide the
observed baryon asymmetry. The top quark takes the role of the Higgsinos.
Focusing on the two-Higgs doublet model, I will discuss details of the
phase transition and consequences for Higgs physics and EDM searches.
}

\section{Introduction}
The baryon asymmetry may have been generated at temperatures much higher
than the electroweak scale, e.g.~through the decay of superheavy right-handed
neutrinos \cite{FY86} or even squarks \cite{H05}. While these proposals seem to
be perfectly viable, they are now, and may remain for some time, very difficult to
test. Electroweak baryogenesis, on the other hand, relies on physics which will
be probed at future colliders, at the first place the LHC.

In recent years, electroweak baryogenesis has mostly been studied in the context
of supersymmetric models, such as the MSSM \cite{CJK00,MSSM} or the NMSSM \cite{NMSSM}.
The CP-violating interactions of the Higgsinos with the expanding bubble walls generate source
terms in the Boltzmann equations which drive the baryogenesis process. 
It is an interesting question if non-supersymmetric models can also 
generate the observed baryon asymmetry. It was the result of  refs.~\cite{BFHS05,FH06}
that this is possible in a general effective field theory approach, where the
SM Higgs sector is augmented by dimension-six operators to induce a first-order
phase transition and to provide additional CP violation. Here we show that a
simple two-Higgs doublet model (2HDM) can explain the baryon asymmetry, without
being in conflict with collider bounds on the Higgs mass or electric dipole moment (EDM)
searches \cite{FHS06}.

\section{The phase transition}
We consider a 2HDM of type II, where the discrete symmetry, introduced to eliminate
flavor violation at the tree-level, is softly broken. The most general potential takes the form 
\begin{eqnarray}
\label{V0}
V_0(H_1,H_2)&=&
-\mu_1^2 H_1^\dagger H_1-\mu_2^2 H_2^\dagger H_2
-\mu_3^2(e^{i\phi} H_1^\dagger H_2+{\rm h.c.})\nonumber\\
&&+\frac{\lambda_1}{2}(H_1^\dagger H_1)^2+\frac{\lambda_2}{2}(H_2^\dagger H_2)^2
+\lambda_3(H_1^\dagger H_1)(H_2^\dagger H_2)\nonumber\\
&&+\lambda_4|H_1^\dagger H_2|^2
+\frac{\lambda_5}{2}\left((H_1^\dagger H_2)^2+{\rm h.c.}\right).
\end{eqnarray}
The model allows us to introduce a single CP-violating phase, which can be
attributed to the soft mass parameter $\mu_3^2e^{i\phi}$. In addition
to the SM Higgs, the 2HDM contains two extra neutral and charged Higgs
particles. In the presence of CP violation, the 3 neutral Higgs states are mixtures, with a scalar 
and pseudoscalar content, and masses $m^2_{H_i}$. 

To reduce the number of parameters, we focus on the case where the extra Higgs
states are heavy and degenerate in mass, i.e.~$m_{H}=m_{H_2}=m_{H_3}=m_{H^\pm}$.
These states will obtain their large masses from order unity self couplings $\lambda_i$.
There is also a lighter SM-like Higgs, with a mass $m_{h}=m_{H_1}$ that has to
fulfill the standard LEP bound of 114 GeV. With this choice large oblique 
corrections are avoided. We also fix the Higgs vev
ratio at $\tan\beta=1$. There is also a lower limit on the charged Higgs mass from
$b\rightarrow s\gamma$  \cite{N04} which will be automatically satisfied in our
parameter region of interest. To $V_0$ we add the one-loop Coleman--Weinberg 
corrections of the heavy Higgs states and the top quark. The Higgs masses are
computed from the full one-loop potential.

To study the electroweak phase transition, we have to compute the finite-temperature
effective potential. We include the one-loop corrections of  the heavy Higgs states, the 
top quark and the gauge bosons. Since the high-temperature approximation is not valid for
the heavy Higgses, we rather use an interpolation to the full one-loop results.

\begin{figure}[t]
\begin{center} 
\epsfig{file=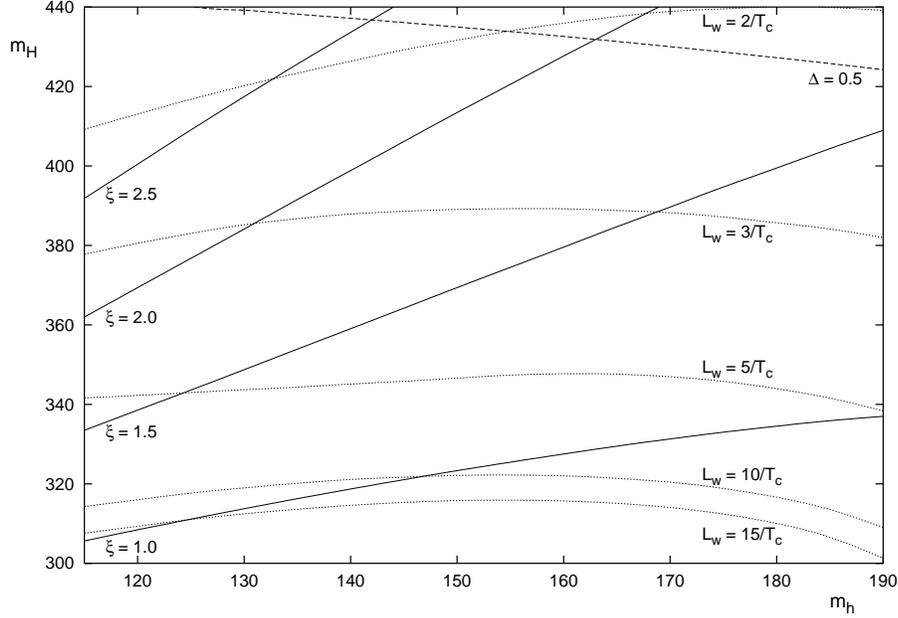,width=82mm,angle=270}
\end{center}
\caption{
Lines of constant $\xi$ and $L_{\rm w}$ in the $m_h$--$m_{H}$ 
plane for $\mu_3^2=10000$ GeV$^2$ and $\phi=0.2$. In addition, the line of
    the relative size of the one-loop corrections
    $\Delta=\max|\delta\lambda_i/\lambda_i|=0.5$ is 
    shown. The Higgs masses are given in units of GeV.
}
\label{fig1}
\end{figure}

The strength of the resulting electroweak phase transition is shown in fig.~\ref{fig1}
\cite{FHS06} for a particular parameter set. Here $\xi=\sqrt{v_{1c}^2+v_{2c}^2}/T_c$
should be larger than 1, so that the phase transition is strong enough to avoid baryon
number washout. As $m_H$ increases, the phase transition becomes stronger.
This somewhat counter-intuitive result is due to the fact that the larger Higgs masses 
come from larger quartic couplings. So this limit actually does not lead to the decoupling 
of the heavy states. At some point perturbation theory will finally break down.
We find that in our example the phase transition becomes sufficiently strong for 
$m_H\gtrsim 300$ GeV, where the size of the zero-temperature one-loop corrections
relative to the tree-level terms is about 15\%, so that perturbation theory is well
under control.  Fig.~\ref{fig1} also shows the line where the corrections become 50\%.
Even for light Higgs masses of $m_h\sim200$ GeV the phase transition can be 
sufficiently strong \cite{FHS06}. So there is a wide parameter range, where the phase
transition is consistent with electroweak baryogenesis.
These results agree with the findings of ref.~\cite{CL96}.

The CP-violating phase $\phi$ has only a minor
impact on the phase transition. We also show the thickness of the expanding
bubble walls, $L_{\rm w}$. As the phase transition gets stronger, the walls
become thinner. In the largest part of the parameter space the walls are thick,
i.e.~$L_{\rm w}T_c\gg1$.

Along the bubble wall  also the relative complex phase between the two Higgs vevs,
$\theta$, changes. In principle one has to numerically solve the field equations of the Higgs fields,
using an algorithm such as the one recently proposed in ref.~\cite{KH06}.
In an approximation, we compute the $\theta$-profile 
by minimizing the thermal potential at $T_c$ with respect to $v_2$ and $\theta$, 
at fixed values of $v_1$, between the symmetric and broken phase. 
For example, we find that for $m_h=150$ GeV, $m_H=350$ GeV, 
$\mu_3^2=10000$ GeV$^2$ and $\phi=0.2$ the phase changes from
$\theta_{\rm sym}=-0.29$ to $\theta_{\rm brk}=-0.06$ \cite{FHS06}.
Later on we will approximate the profiles of $v_i$ and $\theta$ by a kink
ansatz with common wall thickness  $L_{\rm w}$.

\section{The baryon asymmetry}
The CP-violating interactions of particles in the plasma 
with the bubble wall create an excess of left-handed quarks over the corresponding
antiquarks. This excess diffuses into the
symmetric phase, where the left-handed quark density biases the
sphaleron transitions to generate a net baryon asymmetry. Since the bubble walls are thick,
we can apply the WKB formalism \cite{JPT,CJK00,FH06}, and obtain different 
dispersion relations for particles and antiparticles in the space-time dependent background of
the Higgs expectation values. The dispersion relations then lead to force terms
in the transport equations.

In the 2HDM, baryogenesis is driven by top transport. The top quark dispersion
relation to first order in gradients is given by \cite{FH06,PSW1}
\begin{equation} \label{disp}
E=E_0\pm\Delta E=E_0\pm\frac{\theta_t'm_t^2}{2E_0E_{0z}},
\end{equation}
where $E_0=\sqrt{{\bf p}^2+m_t^2}$ and $E_{0z}=\sqrt{p_z^2+m_t^2}$, in terms of the
kinetic momentum, and $\theta_t$ is the phase of the top quark mass. 
The prime denotes the derivative with respect to $z$, the coordinate
perpendicular to the bubble wall. The upper and the lower sign
corresponds to particles and antiparticles, respectively. In ref.~\cite{FH06}
this dispersion relation was derived from the one-particle Dirac equation, and shown to
match the result of the more rigorous Schwinger--Keldysh formalism \cite{PSW1}.
The change in $\theta_t$ along the bubble wall is given
by $\Delta \theta_t= \Delta \theta/(1+\tan^2\beta)$, assuming that
$\tan\beta$ is constant along the wall \cite{HJLS99}. So there is an additional
suppression of  $\Delta \theta_t$ for large $\tan\beta$.

The transport processes in the hot plasma are described by Boltzmann equations.
Top and bottom quarks, and the Higgs bosons are the most important
particle species. The other quark flavors and the leptons can be neglected 
thanks to their small Yukawa couplings.We use a fluid ansatz for the particle 
distribution functions. The CP-violating top dispersion relations enter as force 
terms that source the transport equations. We
take into account $W$-scatterings, the top Yukawa interaction, the strong
sphalerons, the top helicity flips and Higgs number violation with rates
$\Gamma_W$, $\Gamma_y$, $\Gamma_{ss}$, $\Gamma_m$ and $\Gamma_h$,
respectively, where the latter two are only
present in the broken phase. The explicit expressions of the transport equations
are given in ref.~\cite{FHS06} and follow ref.~\cite{FH06}.

\begin{figure}
\begin{center}
   \epsfig{file=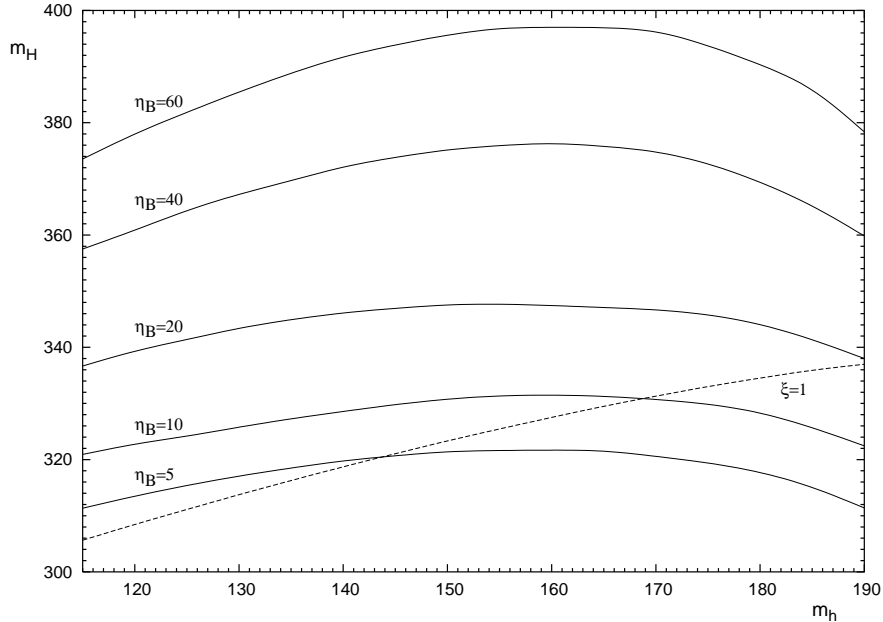,width=82mm,angle=270}
\end{center}
\caption{Contours of constant $\eta_B$ in the $m_h$--$m_H$ plane for 
$\mu^2_3=10000$ GeV$^2$ and $\phi=0.2$. The Higgs masses
  are given in units of GeV and $\eta_B$ in units of $10^{-11}$.
}
\label{fig2}
\end{figure}

In fig.~\ref{fig2} we show the baryon asymmetry $\eta_B=n_B/s$ in the 
$m_h$--$m_H$ plane, fixing again $\mu_3^2=10000$ GeV$^2$ and
$\phi=0.2$. The observational value is \cite{CMB} $\eta_B=(8.7\pm 0.3)\times 10^{-11}$.
For each mass combination we determine all  relevant properties of the phase
transition, such as $\xi$, $L_{\rm w}$, $\theta_{\rm sym}$ and
$\theta_{\rm brk}$ to put them into the transport equations. 
There is only a mild dependence of the baryon asymmetry on the wall velocity, so we
consider only a single value, $v_{\rm w}=0.1$. In addition, the
($\xi$=1)-contour of fig.~\ref{fig1} is also shown for orientation. As we increase
$m_H$, leaving  $m_h$ fixed, the asymmetry becomes larger. 
This behavior results from the $m_t^2\sim\xi^2$ dependence of the top source term.
Accordingly the baryon asymmetry becomes larger for a stronger phase transition.
If we increase $m_h$, leaving 
the heavy Higgs mass fixed, $\eta_B$ becomes smaller and reaches a minimum at
$m_h\approx$ 150--160 GeV, similar to the behavior of $L_{\rm w}$. 
But in general there is only a minor dependence on
the light Higgs mass. 
In this parameter setting it is possible to generate the
observed baryon asymmetry for a heavy Higgs mass between 320 and 330 GeV and a
light Higgs mass up to 160 GeV. Since $\eta_B$ is more or less proportional to
the CP-violating phase $\phi$, the measured value can also be explained for
other  values of the parameters if we adjust $\phi$. 

One can also compute the EDMs of the electron and neutron, 
induced by scalar--pseudoscalar mixing in the neutral Higgs sector.
The dominant contributions are arise from two-loop Barr--Zee type
diagrams. Since there is only a single complex phase in the model, 
we can predict $|d_{e}|$ and $|d_{n}|$ in terms of the baryon asymmetry 
and the Higgs masses. We find that 
$|d_{n}|\gtrsim10^{-27}e~{\rm cm}$. For the smallest allowed values of $m_h$
and $m_H$,  $|d_{n}|$ can slightly exceed the experimental bound of 
$3\times10^{-26}e~{\rm cm}$.
Improving the neutron EDM sensitivity by an order of magnitude would
test a substantial part of the parameter space of the model. The electron EDM is
typically one to two orders of magnitude below the bound of
$1.6\times10^{-27}e~{\rm cm}$. These values are
for $\tan\beta=1$. Extrapolating our results suggests that for $\tan\beta\gtrsim10$,
the 2HDM cannot produce the observed baryon asymmetry without being
in conflict with the EDM constraints. In any case, the 2HDM can explain the
baryon asymmetry for a considerable range of the model parameters. 

\section{Summary}
We have studied electroweak baryogenesis in the 2HDM, focusing on the case of
$\tan\beta=1$ and degenerate extra Higgs states. The phase transition is strengthened
by the loop contributions of the extra Higgs states, provided these are sufficiently
strongly coupled. Taking $\mu_3^2=10000$ GeV$^2$, this happens for a common
heavy Higgs mass $m_H\gtrsim300$~GeV. The mass of the light, SM-like Higgs, 
$m_h$, can be up to 200 GeV, or even larger. The Higgs potential allows us to
introduce of a single CP-violating phase, which induces a varying phase between
the two Higgs vevs along the bubble wall. We compute the CP-violating source term of the top quark
in the WKB approximation and solve the resulting transport equations, using the
formalism of ref.~\cite{FH06}. We find that for typical parameter values
the baryon asymmetry is in the range of the observed value.  Since there is
only a single complex phase in the model, we can predict the electric dipole moments 
 in terms of the baryon asymmetry and the Higgs masses. We find that 
$|d_{n}|\gtrsim10^{-27}e~{\rm cm}$. 
Improving the neutron EDM sensitivity by an order of magnitude would
test a substantial part of the parameter space of the model. The electron EDM is
typically one to two orders of magnitude below the bound. Large values of $\tan\beta$
suppress the baryon asymmetry.

\section*{Acknowledgements}
I wish to thank L.~Fromme and M.~Seniuch for enjoyable collaboration on this project. 
I would also like to thank the conference organizers and acknowledge the financial support
of a Marie Curie grant.

\end{document}